\documentclass[a4paper]{ESASPCS13Style}
\usepackage{epsfig}
\newcommand{\etal}{et\,al.\ }
\newcommand{\logg}{\mbox{$\log g$}}
\newcommand{\Teff}{\mbox{$T_\mathrm{eff}$}}
\newcommand{\hh}{\object{H1504$+$65}}
\newcommand{\alp}{\object{$\alpha$\,Cen}}
\newcommand{\ala}{\object{$\alpha$\,Cen\,A}}
\newcommand{\alb}{\object{$\alpha$\,Cen\,B}}
\newcommand{\alab}{\object{$\alpha$\,Cen\,A and B}}
\newcommand{\pp}{\object{Procyon}}

\newcommand{\hse}{\hbox{}\hspace{1.1mm}\hbox{}}
\newcommand{\hsw}{\hbox{}\hspace{1.3mm}\hbox{}}
\newcommand{\hsed}{\hbox{}\hspace{0.8mm}\hbox{}}
\newcommand{\hses}{\hbox{}\hspace{0.4mm}\hbox{}}
\newcommand{\hsx}{\hbox{}\hspace{1.5mm}\hbox{}}
\begin{document}

\title{Turning cool star X-ray spectra upside down}

\author{K. Werner\inst{1} \and J.\,J.
  Drake\inst{2}} \institute{Institute for Astronomy and Astrophysics, University
  of T\"ubingen, Sand 1, D-72076 T\"ubingen, Germany
  \and Harvard-Smithsonian Center for Astrophysics, MS 3, 60 Garden Street, Cambridge,
    MA 02138, USA }

\maketitle 

\begin{abstract}
H1504+65 is a young white dwarf with an effective temperature of 200\,000\,K and is
the hottest post-AGB star ever analysed with detailed model
atmospheres. {\it Chandra} LETG+HRC-S spectra have revealed the richest
X-ray absorption line spectrum recorded from a stellar photosphere to
date. The line forming regions in this extremely hot photosphere
produce many transitions in absorption that are also observed in
emission in cool star coronae. We have performed a detailed
comparison of {\it Chandra} spectra of H1504+65 with those of Procyon and
$\alpha$\,Cen~A and B. State of the art non-LTE model spectra for the hot
white dwarf have enabled us to identify a wealth of absorption lines
from highly ionised O, Ne and Mg. In turn, these features have allowed us to
identify coronal lines whose origins were hitherto unknown.
\end{abstract}

\begin{figure*}[tbp]
\includegraphics[width=16.5cm]{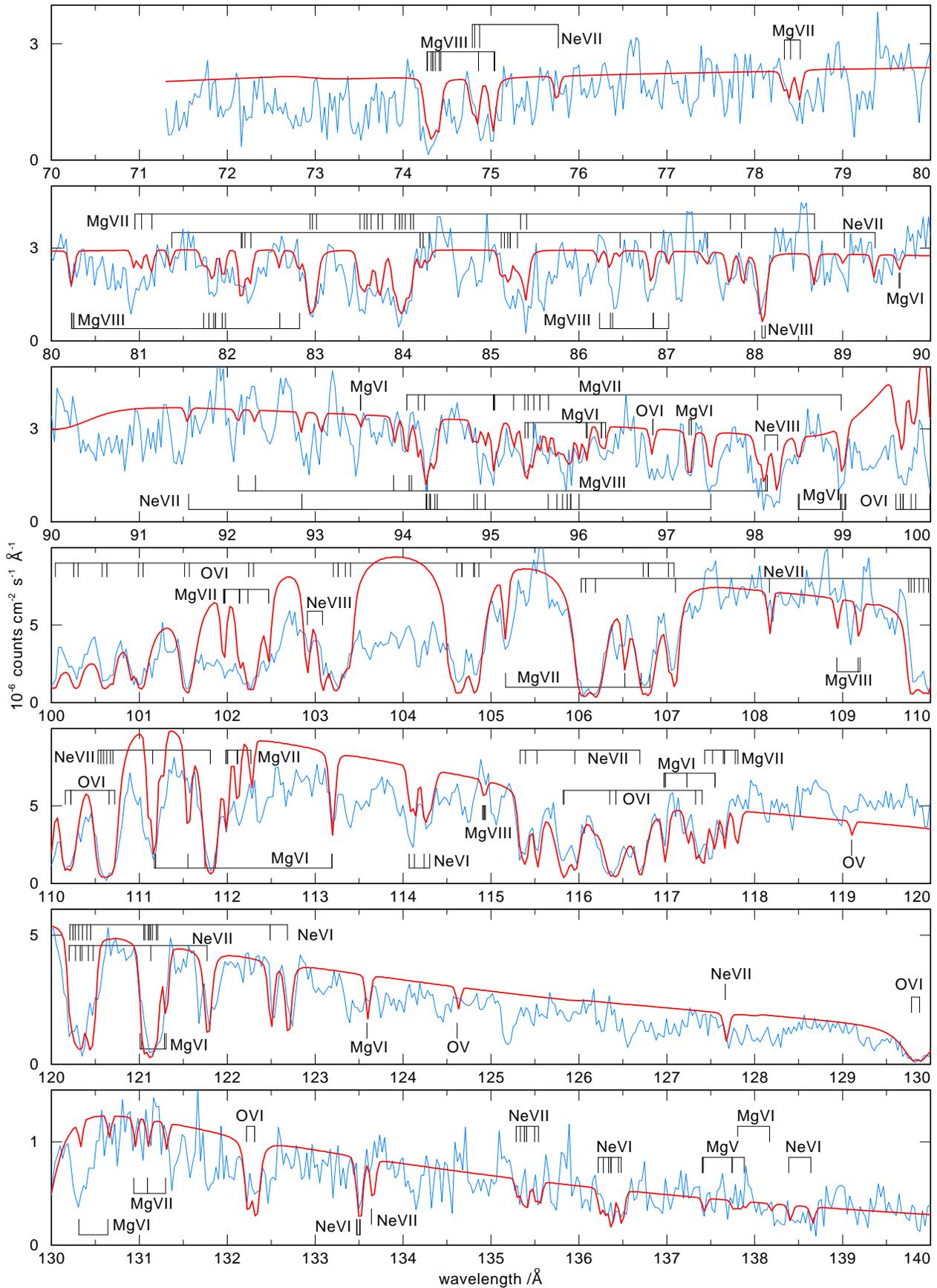} \caption[]{\it The Chandra
            spectrum of \hh\ and a model with
            \Teff=200\,000\,K. 
 } \label{fig_chandra_fit}
\end{figure*}

\begin{figure*}[tbp]
\includegraphics[width=\textwidth]{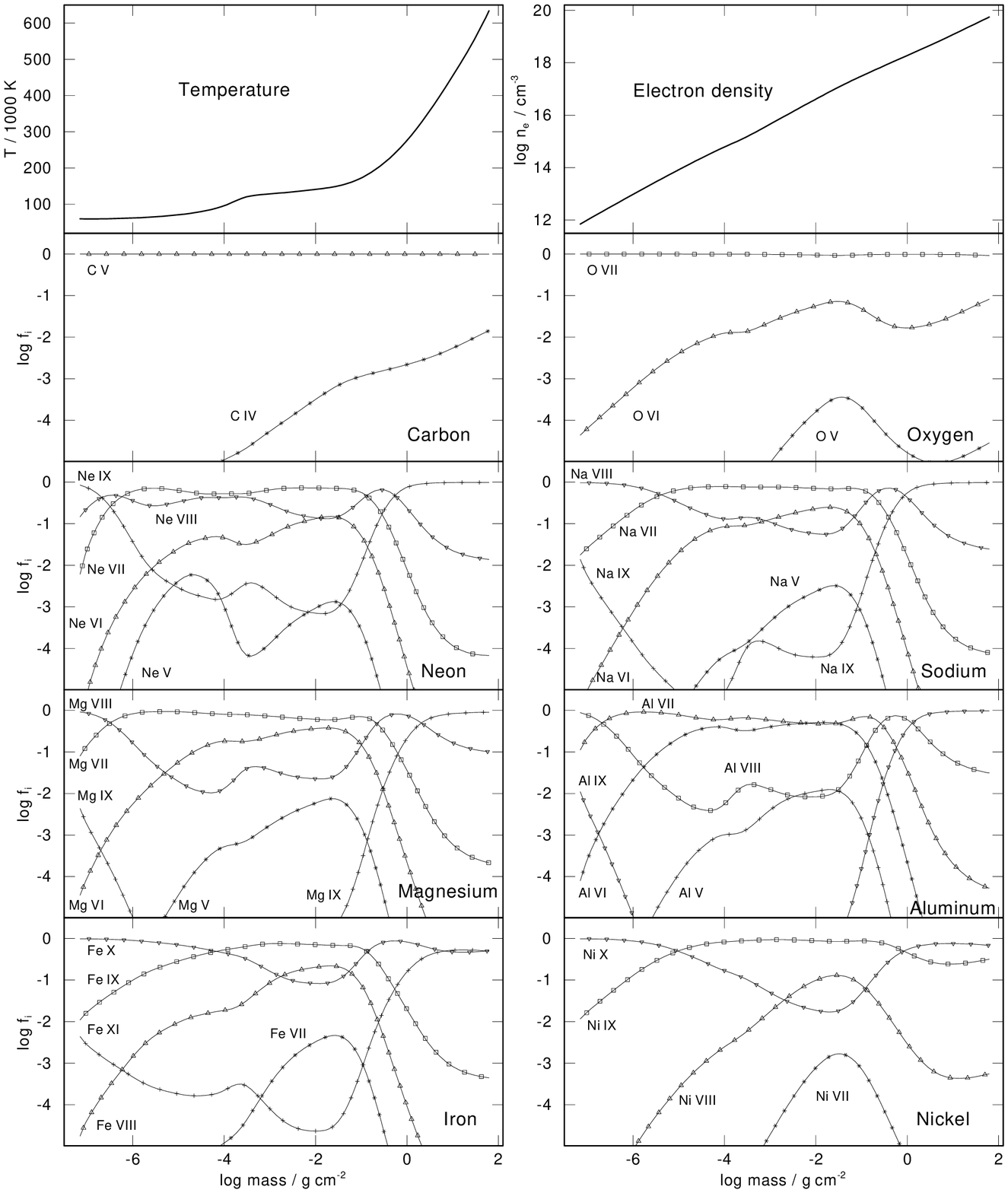} \caption[]{Depth dependence of 
temperature, electron density, and ionization fractions of
            chemical species in a \hh\ model with \Teff=175\,000\,K.} \label{fig_ion}
\end{figure*}

\section{Introduction} 

High-resolution X-ray spectroscopy performed with {\it Chandra} and
XMM-Newton allows very detailed studies of coronae about cool
stars. While many individual emission lines were detected for the
first time in stellar spectra by the Extreme Ultraviolet Explorer
Satellite (EUVE; see, e.g., Drake \etal 1996), the resolving power of
$\lambda/\Delta\lambda\sim 200$ of the EUVE spectrographs was a quite
modest compared with that of present day X-ray observatories.  In
particular, the unprecedented resolution capabilities of the {\it Chandra}
X-ray Observatory Low Energy Transmission Grating Spectrograph (LETG)
in the 30-170~\AA\ range ($\lambda/\Delta\lambda\sim 1000$) that
overlaps with the EUVE Short Wavelength spectrometer (70-170~\AA),
have revealed many more weak spectral lines.  Owing largely to a
historical lack of attention to the 30-170~\AA\ range, a large
fraction of these remain unidentified.  Identification of these
features is desirable because they could be used as a spectroscopic
diagnostics, because they potentially contribute to the flux of
diagnostic lines currently employed, and because they contribute to
the overall plasma radiative loss.  

Two nearby stars that have illuminated the forest of lines in the
30-170~\AA\ range are \alp\ (G2V+K1V) and \pp\ (F5IV).  All three
stars exhibit classical solar-like X-ray emitting coronae.  Indeed,
analogues of the relatively X-ray faint Sun are difficult to observe
because they become unreachable with current instrumentation beyond a
few parsecs, and \alp\ and \pp\ represent the nearest and brightest
coronal sources with solar-like activity.  Only a small fraction of
the multitude of lines between 30-170~\AA\ seen their {\it Chandra} LETG
spectra could be identified based on current radiative loss models
(Raassen \etal\ 2002, 2003).  Drake \etal (in prep.) have estimated
that these models underestimate the true line flux in the range
30-70~\AA\ in these stars by factors of up to 5 or so.

The ``missing lines'' are predominantly transitions involving $n=2$
ground states in abundant elements such as Ne, Mg, Si, S and Ar---the
analogous transitions to the Fe ``L-shell'' lines between $\sim
8$-18~\AA, together with Fe $n=3$ (the ``M-shell'') transitions (Drake
1996, Jordan 1996).  Some of these lines have been identified based on
Electron Beam Ion Trap experiments (Beiersdorfer \etal 1999, Lepson
\etal 2002, 2003).  In the present paper we approach this problem
from a new perspective, namely through a {\it Chandra} observation of the
photosphere of the hottest white dwarf (WD) known, \hh, and its
quantitative analysis by means of detailed non-LTE model atmospheres.

\hh\ has an effective temperature of 200\,000\,K. It belongs to the
PG1159 spectral class, which are hot, hydrogen-deficient (pre-) white
dwarfs. Their surface chemistry (typical abundances: He=33\%, C=48\%,
O=17\%,\\ Ne=2\%, mass fractions) suggests that they exhibit matter from
the helium-buffer layer between the H- and He-burning shells in the
progenitor AGB star (Werner 2001). This is likely because the PG1159
stars have suffered a late He-shell flash, a phenomenon that drives
the fast evolutionary rates of such famous stars like FG\,Sge and
Sakurai's object.  \hh\ is in fact a peculiar member of this class,
because it is also helium-deficient.  Its atmosphere is mainly
composed of carbon and oxygen plus neon and magnesium (C=48\%, O=48\%,
Ne=2\%, O=2\%, mass fractions).  \hh\ is a unique object, considering
its high \Teff\ and chemical surface composition, and we have
speculated that it represents the naked C/O core of a former red giant
(Werner \etal 2004, W04).  

{\it Chandra} LETG+HRC-S spectra from \hh\ have revealed the richest X-ray
absorption line spectrum record\-ed from a stellar photosphere to
date. We have recently performed a detailed analysis of this spectrum
(W04, Fig.~\ref{fig_chandra_fit}) and we use in the paper in hand the photospheric spectrum of
\hh\ together with an appropriate model atmosphere to identify a
number of emission lines in the coronae of \ala, \alb, and
Procyon. The difference in particle densities in the WD photosphere
and in the coronae amounts to many orders of magnitude (roughly
n$_{\rm e}$=$10^{10}$ and $10^{13}-10^{18}$\,cm$^{-3}$, respectively),
however, the temperature in the line forming regions of the WD (up to
300\,000\,K) is comparable to the low-temperature component of
multi-temperature fits to coronae, required to account for the lines
of low-ionisation stages (e.g.\ 630\,000\,K for Procyon; Raassen \etal
2002). As a consequence, numerous lines from \ion{O}{vi},
\ion{Ne}{vi-viii} and \ion{Mg}{vi-ix} are visible in the soft X-ray
spectra of both, the cool star coronae (in emission) and the hot WD
photosphere (in absorption). Lines from higher ionisation stages are
formed in the high-temperature regions of the coronae (T of the order
1--2.5 million K for the stars studied in this paper), hence, their
respective absorption line counterparts cannot be formed in the WD
photosphere. Fig.\,\ref{fig_ion} shows the temperature and particle density
structure of a model for \hh.

In the following, we first introduce briefly the characteristics of
the objects studied here.  We describe our model atmosphere
calculation for the hot WD, concentrating on the atomic data employed.
We then perform a detailed comparison of the absorption and emission
line spectra and suggest a number of new line identifications for the
cool star coronae.

\begin{figure}[tbp]
  \resizebox{\hsize}{!}{\includegraphics{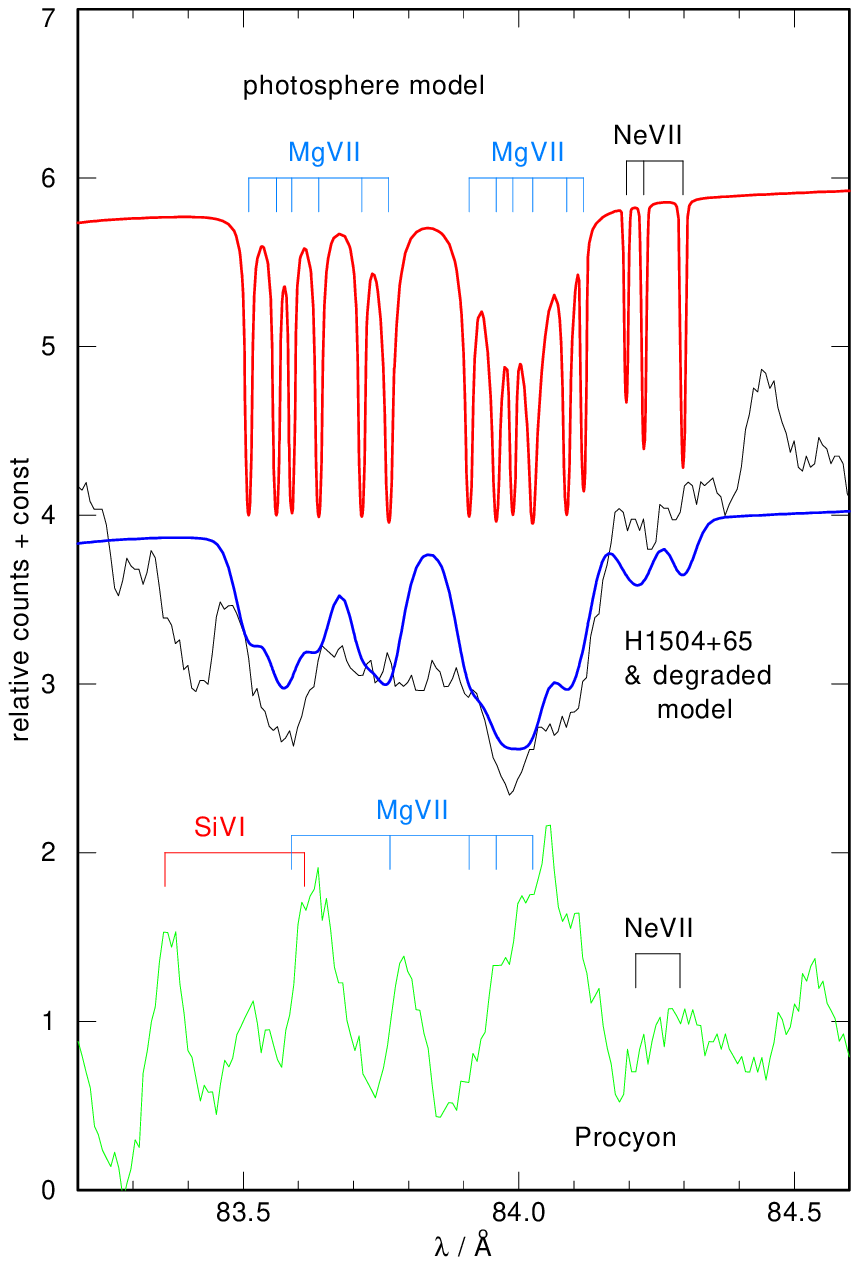}} \caption[]{
  Comparison of {\it Chandra} X-ray spectra of \hh\ and Procyon. Lines from
  \ion{Mg}{vii} and \ion{Ne}{vii} are in absorption in \hh\ and in
  emission in Procyon. Top: photosphere model for \hh\ with line
  identifications for \ion{Mg}{vii} and \ion{Ne}{vii}. Middle:
  Degraded model spectrum (i.e.\ folded with a 0.05\AA\ FWHM Gaussian)
  plotted over \hh\ observation. Bottom: Procyon spectrum with line
  identifications from Raassen \etal (2002). {\it Chandra} spectra were
  smoothed with a 0.03\AA\ boxcar.  } \label{fig_84}
\end{figure}

\begin{figure*}[tbp]
\includegraphics[width=\textwidth]{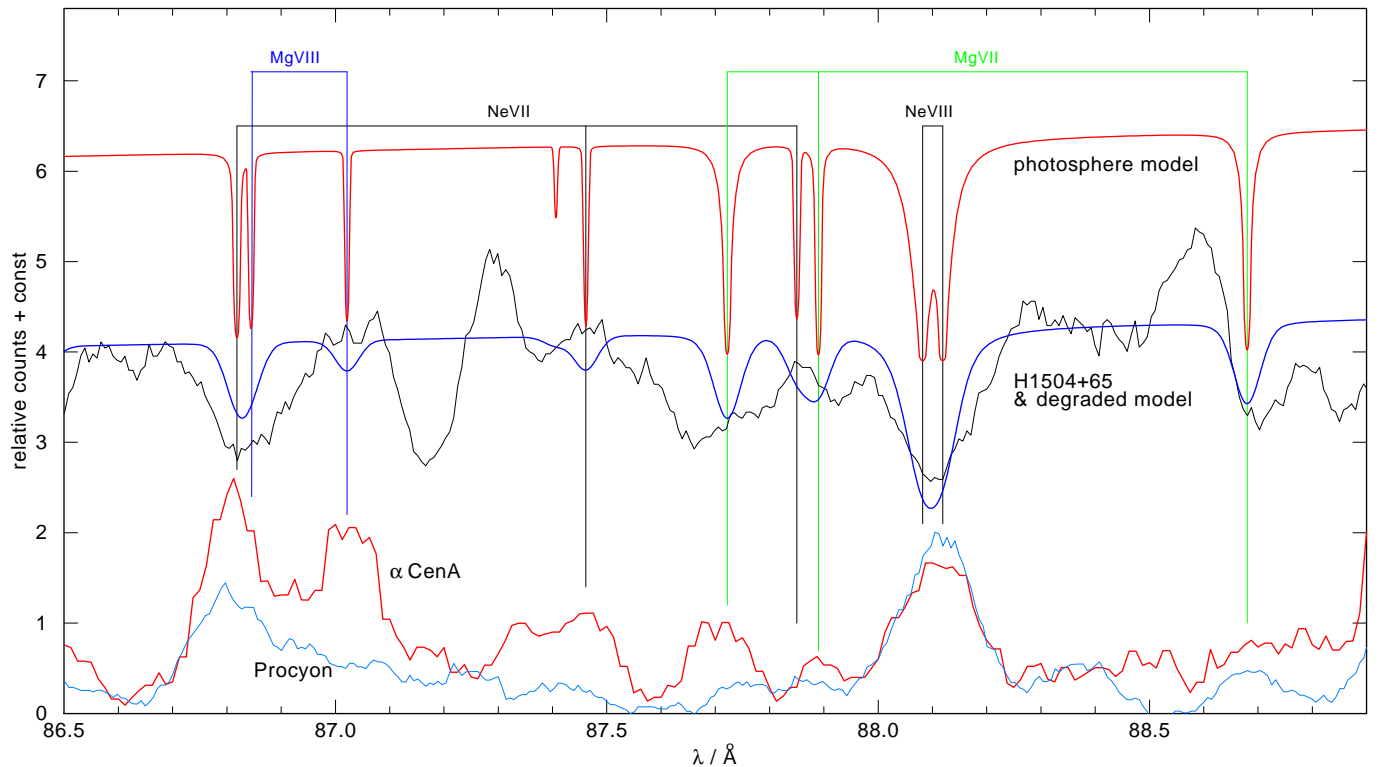} \caption[]{Comparison
  of {\it Chandra} X-ray spectra of \hh\ with Procyon and \ala, similar to
  Fig.\,1. All shown lines from highly ionised Ne and Mg are
  identified for the first time in the cool star corona, except for
  \ion{Mg}{viii}~86.85/87.02\AA\ and the \ion{Ne}{viii} doublet at
  88.1\AA, which were identified by Raassen \etal (2002,
  2003). {\it Chandra} spectra of \hh\ and the coronae were smoothed with
  0.03\AA\ and 0.05\AA\ boxcars, respectively.  } \label{fig_88}
\end{figure*}

\begin{table*}
\caption{ List of X-ray multiplets observed in both the \hh\
photosphere and in the coronae of either \ala\ (``A''), \alb\ (``B''),
or \pp (``P''). In the last column, ``R'' denotes that the line
identification was performed in Raassen \etal (2002, 2003), whereas
``N'' denotes a new identification suggested in this paper. ``R, N''
means that at least one component of the multiplet is newly identified
here. Expressions in brackets denote doubtful cases. The column
``Source'' gives the reference to the level energies of the
transition. After each transition we have marked, if the lower level
is a ground state (``G'') or a metastable state (``M'').
\label{lines_tab}
}
\begin{tabular}{lllr@{\,\,--\,\,}llll}
      \hline
      \hline
      \noalign{\smallskip}
$\lambda$/\AA\ (\hh\ model)     & seen in  & Ion & \multicolumn{2}{c}{Transition} & & Source & Remark \\
      \noalign{\smallskip}
     \hline
      \noalign{\smallskip}
69.41, .47, .57                & A,B,P  & \ion{Mg}{viii} & 2p\hse\,$^2$P$^{\rm o}$  & 3p\hse\,$^2$D\hse       &G & {\sc Nist} & N\\
74.27, .32, .34, .37, .41, .43 & A,B,P  & \ion{Mg}{viii} & 2p$^2$\,$^4$P\hse        & 3d\hse\,$^4$D$^{\rm o}$ &M & {\sc Nist} & N\\
74.78, .81, .87                & A,B    & \ion{Ne}{vii}  & 2p\hsed\,$^3$P$^{\rm o}$ & 4p\,$^3$D\hse           &M & {\sc Nist} & N\\
74.86, 75.03, .04              & A,B,P  & \ion{Mg}{viii} & 2p\hse\,$^2$P$^{\rm o}$  & 3d\hse\,$^2$D\hse       &G & {\sc Nist} & R\\
(78.34), 78.41, 78.52          & A,B,P  & \ion{Mg}{vii}  & 2p$^3$\,$^3$P\hse        & 3p\hse\,$^3$P$^{\rm o}$ &G & Kelly      & N\\
80.23, .25                     & A,B,P  & \ion{Mg}{viii} & 2p$^2$\,$^2$D\hses       & 3d\hse\,$^2$D$^{\rm o}$ &  & {\sc Nist} & R\\
80.95, 81.02, .14              & A,(B,P)& \ion{Mg}{vii}  & 2p$^3$\,$^3$P\hse        & 3p\hse\,$^3$S$^{\rm o}$ &G & Kelly      & N\\
81.37                          & (A),B,P& \ion{Ne}{vii}  & 2p\hsw\,$^1$P$^{\rm o}$  & 4p\,$^1$P\hse           &  & Kelly      & N\\
81.73, .79, .84, .87, .94, .98 & A,B,P  & \ion{Mg}{viii} & 2p$^2$\,$^4$P\hse        & 3s\hse\,$^4$P$^{\rm o}$ &M & {\sc Nist} & N\\
82.17, .20, .27                & A,B,(P)& \ion{Ne}{vii}  & 2p\hse\,$^3$P$^{\rm o}$  & 4d\,$^3$D\hse           &M & Kelly      & N\\
(82.60), .82                   & A,B,P  & \ion{Mg}{viii} & 2p\hse\,$^2$P$^{\rm o}$  & 3s\hse\,$^2$S\hse       &G & {\sc Nist} & R\\
83.51, .56, .59, .64, .71, .76 & A,B,P  & \ion{Mg}{vii}  & 2p$^3$\,$^3$P\hse        & 3d\hse\,$^3$P$^{\rm o}$ &G & Kelly      & R, N\\
83.91, .96, .99, 84.02, .09, .11&A,B,P  & \ion{Mg}{vii}  & 2p$^3$\,$^3$P\hse        & 3d\hse\,$^3$D$^{\rm o}$ &G & Kelly      & R, N\\
(84.19, .23,) .30              & A      & \ion{Ne}{vii}  & 2p\hse\,$^3$P$^{\rm o}$  & 4s\,$^3$S\hse           &M & Bashkin    & R\\
85.41                          & (A,B,P)& \ion{Mg}{vii}  & 2p$^2$\,$^1$D\hses       & 3d\hse\,$^1$F$^{\rm o}$ &M & Kelly      & N\\
86.82                          & A,B,P  & \ion{Ne}{vii}  & 2p$^2$\,$^1$D\hses       & 4d\,$^1$F$^{\rm o}$     &M & Kelly      & N\\ 
86.84, .85, 87.02              & A,B,P  & \ion{Mg}{viii} & 2p$^2$\,$^2$D\hses       & 3s\hse\,$^2$P$^{\rm o}$ &  & {\sc Nist} & R, N\\
87.46                          & A      & \ion{Ne}{vii}  & 2s$^2$\,$^1$S\hsed       & 3s\,$^1$P$^{\rm o}$     &G & {\sc Nist} & N\\
87.72                          & A      & \ion{Mg}{vii}  & 2p$^2$\,$^1$D\hses       & 3d\hse\,$^1$D$^{\rm o}$ &M & Kelly      & N\\
88.08, 88.12                   & A,B,P  & \ion{Ne}{viii} & 2s\hse\,$^2$S\hse        & 3p\,$^2$P$^{\rm o}$     &M & {\sc Nist} & R\\
88.68                          & (A),B,P& \ion{Mg}{vii}  & 2p$^2$\,$^1$S\hsed       & 3d\hse\,$^1$P$^{\rm o}$ &M & Kelly      & N\\
89.64, .65                     & A,(P)  & \ion{Mg}{vi}   & 2p$^3$\,$^2$P$^{\rm o}$  & 4s\hse\,$^2$P           &M & Kelly      & N\\
91.56                          & P      & \ion{Ne}{vii}  & 2p\hsw\,$^1$P$^{\rm o}$  & 4s\,$^1$S\hse           &  & Kelly      & R\\
92.13, .32                     & A,B,P  & \ion{Mg}{viii} & 2p$^2$\,$^2$S\hsed       & 3s\hse\,$^2$P$^{\rm o}$ &  & {\sc Nist} & R, N\\
92.85                          & P      & \ion{Ne}{vii}  & 2p$^2$\,$^1$S\hsed       & 4d\,$^1$P$^{\rm o}$     &M & Kelly      & R\\
(93.89), 94.07, .10, (.27)     & A,B,P  & \ion{Mg}{viii} & 2p\hse\,$^2$P\hse        & 3s\hse\,$^2$P$^{\rm o}$ &  & {\sc Nist} & N\\
94.04, (.17, .24)              & A,B,P  & \ion{Mg}{vii}  & 2p$^3$\,$^5$S$^{\rm o}$  & 3s\hse\,$^5$P\hse       &M & Kelly      & N\\
94.26, .27, .30, .31, .36, .39 & B      & \ion{Ne}{vii}  & 2p\hse\,$^3$P$^{\rm o}$  & 3p\,$^3$P\hse           &M & Bashkin    & N\\
95.03, .04                     & B      & \ion{Mg}{vii}  & 2p$^3$\,$^3$D$^{\rm o}$  & 3s'\,$^3$D              &  & Kelly      & N\\
95.26, .38, .42, .49, .56, .65 & (A,B,P)& \ion{Mg}{vii}  & 2p$^3$\,$^3$P\hse        & 3s\hse\,$^3$P$^{\rm o}$ &G & Kelly      & N\\
(95.38, .42, .48)              & (A,B,P)& \ion{Mg}{vi}   & 2p$^3$\,$^4$S$^{\rm o}$  & 3d\hse\,$^4$P           &G & Kelly      & R\\
95.75, .81, .89, .90, .91, 96.0&A,B,P   & \ion{Ne}{vii}  & 2p\hse\,$^3$P$^{\rm o}$  & 3p\,$^3$D\hse           &M & Bashkin    & N\\
           \noalign{\smallskip}
\hline
     \end{tabular}
\end{table*}

\addtocounter{table}{-1}
\begin{table*}
\caption{continued}
\begin{tabular}{lllr@{\,\,--\,\,}llll}
      \hline
      \hline
      \noalign{\smallskip}
$\lambda$/\AA\ (\hh\ model)     & seen in  & Ion & \multicolumn{2}{c}{Transition} & & Source & Remark \\
      \noalign{\smallskip}
     \hline
      \noalign{\smallskip}
96.08, .09                     & (A,B,P)& \ion{Mg}{vi}   & 2p$^3$\,$^2$P$^{\rm o}$  & 3d''\,$^2$D             &M & Kelly      & N\\
97.50                          & A,B,P  & \ion{Ne}{vii}  & 2s$^2$\,$^1$S\hse        & 3p\,$^1$P$^{\rm o}$     &G & Kelly      & R\\
98.11, .26                     & A,B,P  & \ion{Ne}{viii} & 2p\hse\,$^2$P$^{\rm o}$  & 3d\,$^2$D\hse           &  & {\sc Nist} & R\\
98.50, .51                     & B      & \ion{Mg}{vi}   & 2p$^3$\,$^2$P$^{\rm o}$  & 3d'\,$^2$S              &M & Kelly      & N\\
99.69                          & B      & \ion{O}{vi}    & 2s\hses                  & \hsx  6p                &G & Kelly      & N\\
100.70, .90                    & A      & \ion{Mg}{vi}   & 2p$^3$\,$^2$D$^{\rm o}$  & 3d\,$^2$F               &M & Kelly      & N\\
101.49, .55                    & B      & \ion{Mg}{vi}   & 2p$^3$\,$^2$D$^{\rm o}$  & 3d\,$^2$P               &M & Kelly      & N\\
102.91, 103.08                 & A,B,P  & \ion{Ne}{viii} & 2p\hse\,$^2$P$^{\rm o}$  & 3s\,$^2$S\hse           &  & {\sc Nist} & R\\
103.09                         & (A,B,P)& \ion{Ne}{vii}  & 2p\hsw\,$^1$P$^{\rm o}$  & 3p\,$^1$D\hse           &  & Kelly      & N\\
104.81                         & B,P    & \ion{O}{vi}    & 2s\hses                  & \hsx  5p                &G & Kelly      & R\\
105.17                         & A,(B)  & \ion{Mg}{vii}  & 2p$^3$\,$^1$D$^{\rm o}$  & 3s'\,$^1$D\hse          &  & Kelly      & N\\
106.03, .08, .19               & P      & \ion{Ne}{vii}  & 2p\hse\,$^3$P$^{\rm o}$  & 3d\,$^3$D\hse           &M & Kelly      & R, N\\
(111.10, .16), .26             & A,B,P  & \ion{Ne}{vi}   & 2p\hse\,$^2$P$^{\rm o}$  & 3p\,$^2$D\hse           &G & Kelly      & N\\
111.15                         & (A),B,P& \ion{Ne}{vii}  & 2p$^2$\,$^1$D\hses       & 3d\,$^1$P$^{\rm o}$     &M & Kelly      & N\\
111.55, .75, .86               & B,(A,P)& \ion{Mg}{vi}   & 2p$^3$\,$^4$S$^{\rm o}$  & 3s\hse\,$^4$P           &G & Kelly      & R\\
(115.33), .39, (.52)           & A,B,P  & \ion{Ne}{vii}  & 2p\hse\,$^3$P$^{\rm o}$  & 3s\,$^3$S\hse           &M & Kelly      & R\\
115.82, .83                    & B      & \ion{O}{vi}    & 2s\hses                  & \hsx  4p                &G & Kelly      & R\\
115.96                         & B      & \ion{Ne}{vii}  & 2p$^2$\,$^1$D\hses       & 3d\,$^1$D$^{\rm o}$     &M & Kelly      & N\\
(116.35), .42                  & B      & \ion{O}{vi}    & 2p                       & \hsx  5d                &  & Kelly      & N\\
116.69                         & B      & \ion{Ne}{vii}  & 2p\hsw\,$^1$P$^{\rm o}$  & 3d\,$^1$D\hse           &  & Kelly      & R\\
116.97, 117.22                 & A      & \ion{Mg}{vi}   & 2p$^3$\,$^2$P$^{\rm o}$  & 3s'\,$^2$D              &M & Kelly      & N\\
(117.33), .40                  & B      & \ion{O}{vi}    & 2p                       & \hsx  5s                &  & Kelly      & N\\
(117.43), .66, (.78)           & P      & \ion{Mg}{vii}  & 2p$^3$\,$^3$S$^{\rm o}$  & 3s\hse\,$^3$P\hse       &  & Kelly      & N\\
(117.52), .64, (.81)           & P      & \ion{Mg}{vii}  & 2p$^3$\,$^3$P$^{\rm o}$  & 3p\hse\,$^3$P\hse       &  & Kelly      & N\\
120.20, .27, .33, .35, .42, .48& P      & \ion{Ne}{vii}  & 2p$^2$\,$^3$P\hse        & 3s\,$^3$P$^{\rm o}$     &  & Kelly      & N\\
122.49, .69                    & B,P    & \ion{Ne}{vi}   & 2p\hse\,$^2$P$^{\rm o}$  & 3d\,$^2$D\hse           &G & Kelly      & R, N\\
123.59                         & P      & \ion{Mg}{vi}   & 2p$^4$\,$^2$D\hse        & 3s$^{\rm iv}$\,$^2$D$^{\rm o}$& &Kelly & N\\
127.67                         & B,P    & \ion{Ne}{vii}  & 2p\hsw\,$^1$P$^{\rm o}$  & 3s\,$^1$S\hse           &  & Kelly      & R\\
129.78, .87                    & A,B,P  & \ion{O}{vi}    & 2p                       & \hsx  4d                &  & Kelly      & R\\
130.31, .64                    & B      & \ion{Mg}{vi}   & 2p$^4$\,$^2$P\hse        & 3s$^{\rm v}$\,$^2$P$^{\rm o}$& &Kelly  & N\\
130.94, 131.09, .30            & A,B,P  & \ion{Mg}{vii}  & 2p$^3$\,$^3$S$^{\rm o}$  & 3p\hse\,$^3$P\hse       &  & Kelly      & N\\
132.22, .31                    & A,B    & \ion{O}{vi}    & 2p                       & \hsx  4s                &  & Kelly      & N\\
150.09, .12                    & B,P    & \ion{O}{vi}    & 2s\hses                  & \hsx  3p                &G & Kelly      & R\\
           \noalign{\smallskip}
\hline
     \end{tabular}
\end{table*}

\section{Observations} 

\hh\ was observed with the {\it Chandra} LETG+HRC-S on September 27, 2000, with an
integration time of approximately 25\,ks. Flux was detected in the range
60\AA--160\AA. The spectrum is that of a hot photosphere,
characterized by a continuum with a large number of absorption lines
from highly ionized species: \ion{O}{v-vi}, \ion{Ne}{vi-viii}, and
\ion{Mg}{v-viii}.  It rolls off at long wavelengths due to ISM
absorption. The maximum flux is detected near 110\AA. Between 105\AA\
and 100\AA\ the flux drops because of photospheric absorption from the
\ion{O}{vi} edge caused by the first excited atomic level. The edge is
not sharp because of a converging line series and pressure
ionization. Below 100\AA\ the flux decreases, representing the Wien
tail of the photospheric flux distribution. The complete spectrum with
detailed line identifications was presented in W04.

The \alab\ observation has been described in detail by Raassen \etal\
(2002) and we describe it here only in brief.  \alp\ was observed
with the LETG+HRC-S on December 25, 1999 with an exposure time of 81.5\,ks, 
including dead time corrections to account for telemetry
saturation during intervals of high background.  The observation was
designed such that the two stars were maximally separated in the
cross-dispersion axis, with the dispersion axis positioned nearly
perpendicular to the axis of the binary.  At the time of the
observation, the stars were separated by $16\arcsec$ on the sky. The
spectra were extracted with the standard CIAO bow-tie region, though
the central two background regions interfered with the stellar spectra
and only the outer regions were used for background subtraction.

The two Procyon observations studied here were obtained with the 
LETG+HRC-S as part of the {\it Chandra} on-orbit calibration
programme and Emission Line Project.  The observations were executed
contiguously beginning on November 6, 1999 at
21:11:32 UT. The second observation began on 1999 November 16:59:48
UT.  The effective exposure times for these observations were 69,643s
and 69,729s, respectively, including dead time corrections.

Reduction of the HRC-S event lists for all the observations was
initially based on standard pipeline products.  Events were further
filtered in pulse height in order to remove background events. The
final reduced first order spectra were co-added in order to maximise
the signal.  In the case of Procyon, we also co-added the two separate
observations.

\section{Photospheric model for \hh}

We use here a photospheric spectrum from a line blanketed non-LTE
model atmosphere constructed for \hh\ by W04. Model parameters are:
\Teff=200\,000\,K, \logg=8 [cm/s$^2$], and C=48\%, O=48\%, Ne=2\%,
O=2\%, (mass fractions). Details of model assumptions and calculations
can be found in that reference and we restrict ourselves here to those
characteristics which are of immediate relevance in our context.  This
primarily concerns the NLTE model atoms for neon and magnesium. They
comprise 88 and 122 NLTE levels, connected with 312 and 310 radiative
line transitions, respectively, in the ionization stages
\ion{}{iv-ix}. The final synthetic spectrum was computed considering
fine structure splitting of levels and multiplets assuming relative
LTE populations for levels within a particular term. We have tried to
use the best available data for level energies and line wavelengths,
compiling them from several sources. For the lines discussed here
(Table~\,\ref{lines_tab}), we used the following databases:

\noindent
(i) National Inst.\ of Standards and Technology (NIST)\footnote{http://physics.nist.gov/},\\
(ii) {\sc Chianti} database (Young \etal 2003)\footnote{http://wwwsolar.nrl.navy.mil/chianti.html},\\
(iii) Kelly Atomic Line Database\footnote{http://cfa-www.harvard.edu/amdata/ampdata/kelly/\\kelly.html}.

\noindent
However, in order to assemble the complete model atoms, other sources were essential, too:

\noindent
(iv) Opacity Project (OP, Seaton \etal 1994) TOPbase\footnote{http://legacy.gsfc.nasa.gov/topbase/home.html},\\ 
(v) University of Kentucky Atomic Line List\footnote{http://www.pa.uky.edu/$^\sim$peter/atomic/}.

\section{Comparison with \alab, and \pp} 

We have performed a detailed comparison of the \hh\ photospheric
absorption line spectrum with the coronal emission line spectra of
\ala, \alb, and \pp. We have also used the model spectrum of \hh\ for
this purpose. It turns out that not all lines predicted by the model,
particularly the weaker ones, are readily identified in \hh, which is
at least in part due to the S/N of the {\it Chandra} spectrum. Another
reason is heavy blending by lines from iron group elements, which are
not considered in the model used here. It was shown that
identification of weak lines suffers from iron and nickel line blends,
which is a problem because the accurate positions of the majority of
lines from Fe-group elements in the soft X-ray domain is unknown
(W04).  The use of our synthetic spectrum in addition to the \hh\
spectrum helps considerably to identify lines in the coronal spectra.

Table~\,\ref{lines_tab} summarizes the results of our
comparison. Lines from 65 multiplets of \ion{O}{vi},
\ion{Ne}{vi-viii}, and \ion{Mg}{vi-viii} are identified in both, \hh\
(or its model) and in at least one of the considered coronae. Many of
these were already identified by Raassen \etal (2002, 2003), but the
majority represents new identifications. Many, but not all, of the
tabulated lines have lower levels which are either ionic ground states
or metastable states (labeled G or M, respectively).  As an example
how the spectra compare, we show in Fig.\,\ref{fig_84} the spectra of
Procyon and \hh\ in a wavelength region where a bunch of lines from
two \ion{Mg}{vii} and one \ion{Ne}{vii} multiplet is located. All
three multiplets, or at least some components of them, were identified by
Raassen \etal (2002) in Procyon. They are also clearly seen as
absorption features in the \hh\ spectrum. Over this, we have plotted
the model spectrum, degraded to the {\it Chandra} spectral resolution, which
can qualitatively reproduce the observed line features. Placed at the
top of this Figure we show the original, non-degraded model spectrum,
showing the diverse structure of the multiplets, whose components are
not entirely resolved in {\it Chandra} spectra, neither of \hh\ nor of \pp.

Fig.\,\ref{fig_88} shows a detail from the spectra of \pp\ and \ala\
compared to \hh\ in another wavelength interval. It displays some new
line identifications in the coronal spectra, see for example the 87.46\AA\
resonance line of \ion{Ne}{vii} in \ala. The strongest emissions in
\ala\ stem from two \ion{Ne}{viii} and \ion{Mg}{viii} doublets,
identified already in Raassen \etal (2003). But note that the
\ion{Mg}{viii}~86.84\AA\ component is blended with the possibly
stronger, newly identified \ion{Ne}{vii}~86.82\AA\ line.

Some of the lines newly identified lines do blend with other lines used for
coronal diagnostics.  The emissivity of the \ion{Fe}{viii} lines at
130.94\AA\ and 132.24\AA\ in Procyon was computed by Raassen \etal
(2002) using a three-temperature model. They stress that these line
strengths are strongly underestimated, by factors 6 and 4 compared to
the observation. The result of their differential emission measure
(DEM) model underestimates the emissivity even more (factors 9 and
6). This can at least partially be explained by the fact that two
components of a \ion{Mg}{vii} triplet (at 130.94\AA\ and 131.30\AA)
can contribute to the \ion{Fe}{viii} line emissivities. A similar
explanation may hold for the too-weak \ion{Fe}{ix}~105.20\AA\ line in
the model. It is blended with a \ion{Mg}{vii} singlet at 105.17\AA.

Another example is the \ion{Mg}{viii}~74.86\AA\ line observed in \ala\
and \alb. Raassen \etal (2003) find that the line fluxes from their
models are too small by about 40\%. We think that the missing flux is
contributed by a blend with a new neon line located at almost the same
wavelength, \ion{Ne}{vii}~74.87\AA. Detailed emission measure
modeling, which is beyond the scope of this paper, is needed to
quantify these suggestions.  Other blends with previously identified
emission lines in the coronae of Procyon and $\alpha$\,Cen are
indicated in Table~\ref{lines_tab}.

\section{Summary}

We have performed a detailed comparison of {\it Chandra} soft X-ray spectra
from the photosphere of the hottest known white dwarf, \hh, with the
corona spectra of \ala, \alb, and \pp. With the help of a detailed
model spectrum for \hh\ we have found that a large number of lines
from multiplets of O, Ne, and Mg are present in both the photospheric
absorption line spectrum and the coronal emission line spectra. In the
coronal spectra we have newly identified lines from about 40
multiplets of \ion{O}{vi}, \ion{Ne}{vi-vii}, and
\ion{Mg}{vi-viii}. Some of these lines are blends with previously
known lines, which are in use for diagnostic purposes, hence, their
contribution to the line flux must be considered in detailed spectral
analyses.

A more complete version of this paper will be published in Astronomy \& Astrophysics.

\begin{acknowledgements}
Analysis of X-ray data in T\"ubingen is supported by the DLR under
grant 50\,OR\,0201. JJD was supported by NASA contract NAS8-39073 to the {\it
  Chandra X-ray Center}. 
\end{acknowledgements}

\end{document}